\documentclass[12pt,a4paper]{JHEP3}
\usepackage{rotating}
\usepackage{graphics}
\usepackage{graphicx}
\usepackage{epsfig}
\usepackage{amsmath}
\usepackage{quotes}

\usepackage{amssymb}
\usepackage{cite}

\def\AdS5S5{\mathrm{AdS}_5\times \mathrm{S}^5}

\newcommand{\Tr}{{\textrm{Tr}}}

\newcommand{\be}{\begin{equation}}
\newcommand{\ee}{\end{equation}}
\newcommand{\bea}{\begin{align}}
\newcommand{\eea}{\end{align}}

\title{Algebraic Curves for Factorized String Solutions}

\author{
Amit~Dekel
\\
Department of Physics and Astronomy, Uppsala University\\
SE-75108 Uppsala, Sweden\\
\email{amit.dekel@physics.uu.se}}

\abstract{
We show how to construct an algebraic curve for factorized string solution in the context of the AdS/CFT correspondence.
We define factorized solutions to be solutions where the flat-connection becomes independent of one of the worldsheet variables by a similarity transformation with a matrix $S$ satisfying $S^{-1}d S=const$.
Using the factorization property we construct a well defined Lax operator and an associated algebraic curve.
The construction procedure is local and does not require the introduction of a monodromy matrix.
The procedure can be applied for string solutions with any boundary conditions.
We study the properties of the curve and give several examples for the application of the procedure.
}

\preprint{UUITP-04/13}

\begin{document}

\newpage

\section{Introduction}

In this paper we study the algebraic curve of a special class of string solutions in the context of the AdS/CFT correspondence \cite{Maldacena:1997re,Gubser:1998bc,Witten:1998qj} (see \cite{Aharony:1999ti} for review).
String theory on an $\mathrm{AdS}_5\times \mathrm{S}^5$ background is known to be integrable \cite{Bena:2003wd}
by having an infinite set of conserved charges which are encoded in the monodromy matrix (see \cite{Beisert:2010jr} for a review).
However, it is still a very hard problem to find string solutions in the $\mathrm{AdS}_5\times \mathrm{S}^5$ background.

A study of the spectral problem on the string theory side of the AdS/CFT correspondence revealed that closed string solutions are classified and encoded by algebraic curves \cite{Kazakov:2004qf,Kazakov:2004nh,Beisert:2004ag,Beisert:2005bm}.
The construction of the algebraic curve for a closed string solution is based on the existence of a monodromy matrix with the integration defined along a non-contractible loop circling the cylinder defined by the closed string worldsheet.
It was shown in \cite{Janik:2012ws} that some string solutions where non-trivial monodromies do not exist (e.g open string solutions) may also admit a non-trivial algebraic curve (see also \cite{Ryang:2012uf} for more examples).
The construction of \cite{Janik:2012ws} is based on the existence of a Lax operator and its analytical properties, which may be constructed directly with no reference to a monodromy matrix.

In this paper we consider a special class of string solutions where the construction of the Lax operator is significantly simplified.
We show how one can easily extract the algebraic curve for this class of solutions and analyze the properties of these curves.
This class of solutions is given by solutions for which the flat-connection is factorized in terms of the worldsheet coordinates in the sense that the time evolution of the flat-connection is given by a similarity transformation, $A(\tau,\sigma)=S^{-1}(\tau)A_0(\sigma)S(\tau)$ with $S^{-1}d S=const.$
We refer to such solutions as \textit{factorized solutions}.
Factorized string solutions were also considered in \cite{Arutyunov:2003za} where similar ideas concerning the construction of the Lax operators were presented, however not in the context of the algebraic curve.
This class of solutions contains many of the well known string solutions in AdS background such as the null cusp \cite{Kruczenski:2002fb}, quark-antiquark potential \cite{Maldacena:1998im,Rey:1998ik,Chu:2009qt}, two point function \cite{Janik:2010gc}, folded spinning string \cite{Gubser:2002tv} etc. (actually all the string solutions which were considered in \cite{Janik:2012ws} and \cite{Ryang:2012uf} fall into this class).

The paper is organized as follows.
In section \ref{sec:fac_sol_4_group_man} we introduce the factorized solutions for sigma-models on group manifolds and construct the Lax operator and the algebraic curve.
In section \ref{sec:prop_4_group_man} we discuss the properties of the curve for strings on an AdS$_3$ background.
We also discuss the reconstruction of the string solutions from the curve.
In section \ref{sec:AdS5S5Curves} we give a short discussion on the construction of the curve in the $\mathrm{AdS}_5$ background.
In section \ref{sec:examp_4_group_man} we give several examples of applications for our procedure.
Some of these examples are new while some were already discussed in \cite{Janik:2012ws} and \cite{Ryang:2012uf}, but we present them for completeness as well as comparison with the results that our procedure yields.
We also give an example of a string solution in AdS$_5$.
In section \ref{sec:fac_sol_4_coset_man} we give a short discussion on the generalization of the method for string solutions on super-coset backgrounds.
We end with a discussion in section \ref{sec:discussion}.

\section{Factorized String Solutions for Sigma-Models on Groups Manifolds}\label{sec:fac_sol_4_group_man}

In this section we consider string sigma-models on a group manifold $G$, namely
\begin{align}
S=\int d^2\sigma \Tr (j\wedge \ast j),
\end{align}
where $j=g^{-1}d g$ is the Maurer-Cartan one-form taking values in the algebra $\mathfrak{g}$ of $G$, and $g\in G$.
We consider solutions with the following factorization with respect to the worldsheet coordinates
\begin{align}
j(\sigma,\tau)=S^{-1}(\tau)j_0(\sigma)S(\tau), \quad S(\tau)\in G,\quad j_{0}(\sigma)\in \mathfrak{g},
\end{align}
where $j_S^R\equiv d S S^{-1}=const.$ and $j_S^L\equiv S^{-1}d S =const.$ (note that $j_{S,\sigma}^{L/R}=0$).
Note that this equation does not imply that $j_0(\sigma)$ has a vanishing $\tau$ or $\sigma$ components\footnote{Note that the subscript $'0'$ implies $\tau=0$ and not the $\tau$ component.}.

The factorization of the MC one-form implies that the flat-connection, $A=\frac{1}{1-z^2}\left(j-i z \ast j\right)$, inherits the same factorization properties\footnote{The presence of the $i$ factor is due to the use of a Euclidean worldsheet signature.
The worldsheet signature does not affect our analysis and the final results are the same for a worldsheet with Minkowskian signature.} ($z\in \mathbb{C}$ is the spectral parameter).
The flatness equation of a factorized flat-connection takes the form
\begin{align}\label{eq:Flatness_eq}
\partial_\sigma A_{0,\tau}(\sigma)&=[A_{0,\tau}(\sigma)-j_S^R,A_{0,\sigma}(\sigma)].
\end{align}
This equation suggests we should define a new variable
\begin{align}\label{eq:LaxOp}
L(\tau,\sigma)\equiv A_\tau(\tau,\sigma) - j_S^L,
\end{align}
taking values in the algebra.
By (\ref{eq:Flatness_eq}) $L$ satisfies
\begin{align}\label{eq:Lax1}
\partial_\sigma L&=[L,A_{\sigma}].
\end{align}
By taking the $\tau$ derivative we find
\begin{align}\label{eq:Lax2}
\partial_\tau L&=[L,j_S^L]=[L,j_S^L+L]=[L,A_\tau].
\end{align}
From now on we denote $j_S^L$ by $j_S$.
Thus, $L$ is a Lax operator satisfying the Lax equations
(note that the Lax equations are linear and that $L^m$ also satisfies them for a non-negative integer $m$).
The Lax equations imply that the algebraic curve equation is given by \cite{Janik:2012ws}
\begin{align}\label{eq:AlgCurveEq}
\det(L-y 1_{n\times n})=0.
\end{align}
The Lax operator depends on the spectral parameter in a simple way by construction.
The characteristic polynomial equation is given by 
\begin{align}
\det(L-y 1_{n\times n})\propto \frac{1}{(1-z^2)^n}\sum_{i=1}^{2n} c_i(y) z^{i}
\end{align}
 where $c_i(y)$ are polynomials in $y$ up the power $n$.
Note that if $A$ is traceless, 
\begin{align}
\partial_\alpha (\det(L-y ))\propto \Tr\left((L-y )^{-1}\partial_\alpha(L-y )\right)=\Tr\left((L-y )^{-1}[L-y,A_\alpha)]\right)=0,
\end{align}
which is the case for the backgrounds we will be interested in, namely  AdS$_3$ and AdS$_5$.

Writing the flatness equation in terms of the currents gives
\begin{align}
\left(\partial_\sigma j_\tau-i z \partial_\sigma j_\sigma\right)
&=\frac{1}{1-z^2}[\left(j_\tau-i z j_\sigma\right)-(1-z^2)j_S,\left(j_\sigma+i z j_\tau\right)].
\end{align}
Expanding the above equation in $z$ we find the "factorized" Maurer-Cartan equations and equations of motion
\begin{align}
\partial_\sigma j_\tau
&=[j_\tau-j_S,j_\sigma],\nonumber\\
\partial_\sigma j_\sigma
&=[j_S,j_\tau],
\end{align}
respectively.
Using these equations it is easy to check that the energy-momentum tensor components are constant
\begin{align}
\partial_\sigma T_{\tau\sigma}&=\partial_\sigma \Tr(j_\tau j_\sigma)
=
 \Tr([j_\tau-j_S,j_\sigma] j_\sigma)
+\Tr(j_\tau [j_S,j_\tau])=0
,\nonumber\\
\partial_\sigma T_{\tau\tau}&=\partial_\sigma \Tr(j_\tau j_\tau-j_\sigma j_\sigma)
=2\partial_\sigma \Tr(j_\tau [j_\tau-j_S,j_\sigma]-j_\sigma [j_S,j_\tau])=0.
\end{align}
The vanishing of the $\tau$ derivative is obvious.

Note that we could have also started with a factorization of the solution with respect to $\sigma$ instead of $\tau$.
In this case the Lax operator is given by $L=A_\sigma-j_S$ (where $j_{S}=(j_S)_\sigma$).
For example, this is the case for the algebraic curve for $\left< W(C)\Tr(Z^J) \right>$ as we shall see in section \ref{sec:WrtZ^Jexample}.
Furthermore, the factorization could also be with respect to some combination of $\tau$ and $\sigma$, and the generalization of the Lax operator in that case should be obvious.

\section{The AdS$_3$ Case}\label{sec:prop_4_group_man}

In this section we consider the case of $G=\mathrm{SL}(2,R)$ in the $\mathbf{2}$ representation which gives a sigma model on $\mathrm{AdS}_3$ (the analysis of the real form $G=\mathrm{SU}(2)$ which gives $\mathrm{S}^3$ is identical).
First we write the curve in terms of the currents and then we discuss its properties.

\subsection{The Algebraic Curve}
We start by noting that $\det(a)=-\frac{1}{2}\Tr(a^2)$ for $a\in \mathrm{sl}(2)$, since $a$ is traceless.
Thus the algebraic curve equation reduces to
\begin{align}\label{eq:AdS3AlgCurve}
y^2=-\det(L).
\end{align}
Explicitly we have
\begin{align}
\det(L)=-\frac{1}{2}\Tr(L^2)
&=
-\frac{1}{2}\Tr((\frac{1}{1-z^2}(j_\tau-i z j_\sigma)-j_S)^2)
=-\frac{1}{2}\frac{1}{(1-z^2)^2}\Tr((j_\tau-j_S)^2\nonumber\\
&
-2 i z(j_\tau-j_S)j_\sigma
+z^2(2 (j_\tau-j_S) j_S-j_\sigma^2)
-2 i z^3 j_\sigma j_S
+z^4 j_S^2)\nonumber\\
&
\equiv -\frac{1}{2}\frac{1}{(1-z^2)^2}\sum_{i=0}^{4}c_i z^i.
\end{align}
We define the polynomial $g(z)=\sum_{i=0}^{4}c_i z^i$ which characterizes the algebraic curve (where we used Euclidean worldsheet metric).
We may rescale $L$ or $y$ such that we get an algebraic curve equation $y^2=g(z)$.
Such a curve gives at most a genus-1 Riemann surface.
We write the coefficients $c_i$ explicitly in terms of traces of the currents
\begin{align}\label{eq:AdS3coeff}
c_0&=\Tr((j_\tau-j_S)^2),\nonumber\\
c_1&=-2 i \Tr((j_\tau-j_S)j_\sigma),\nonumber\\
c_2&=\Tr(2 (j_\tau-j_S) j_S-j_\sigma^2),\nonumber\\
c_3&=-2 i \Tr(j_\sigma j_S),\nonumber\\
c_4&=\Tr(j_S^2).
\end{align}

\subsection{Properties of the Algebraic Curve}

Next, we collect a number of properties of the AdS$_3$ algebraic curve.
Some of the statements presented in this section will be proved in section \ref{sec:examp_4_group_man}.

\subsubsection{The Virasoro Constraints}\label{sec:Virasoro}

Given (\ref{eq:AdS3coeff}) we can easily read the energy-momentum tensor components from the curve's coefficients
\begin{align}
T_{\tau\sigma}&=i\frac{c_1+c_3}{2},\nonumber\\
T_{\tau\tau}&=c_0+c_2+c_4.
\end{align}
These relations imply that the Virasoro constraints are satisfied only if the curve has the form
\begin{align}
g(z)=(1-z^2)f(z),
\end{align}
where $f(z)$ is a second order polynomial.
We will show later that indeed, the algebraic curves of the null cusp and its $\mathrm{SO}(2)$ worldsheet rotations, the circular Wilson loop, the folded spinning string and the $q\bar q$ potential all take this form (this fact was also observed and noted in \cite{Ryang:2012uf} for the considered examples).

\subsubsection{Symmetries}

The algebraic curve is invariant under the global symmetries of the action, that is $\mathrm{SO}(d,2)$ conformal transformation for strings on $\mathrm{AdS}_{d+1}$ background.
It is also invariant under translations along the worldsheet.

In our analysis we use the worldsheet conformal gauge, in which the action and equations of motion are invariant under an $\mathrm{SO}(2)$ rotation of the worldsheet coordinates.
Explicitly such transformations are given by $\xi=(\tau,\sigma)\rightarrow \xi'=(\tau',\sigma')=(\tau\cos\theta+\sigma\sin\theta,\tau\sin\theta-\sigma\cos\theta)$.
Under such a transformation the Lax operator transforms as $L(\xi)\rightarrow \cos\theta L'(\xi')+\sin\theta A_{\sigma'}(\xi')$,
so our algebraic curve with respect to $\tau$ is not invariant under such transformations.
Also, the factorization of the solution with respect to $\tau$ might be lost in general, and the solution is then factorized with respect to $\tau'$.
Obviously the same curve still exists but now with respect to $\tau'$ factorization.

Under a Wick rotation where $\tau\rightarrow i\tau$ the curve transforms as $g(z)\rightarrow - g(z)$, and under a rescaling of the worldsheet variables $\xi\rightarrow \xi'=\alpha \xi$ the curve transforms as $g(z)\rightarrow \alpha^2 g(z)$.

\subsubsection{Complete Factorization}
Solutions which are completely factorized, that is $j(\sigma,\tau)=S^{-1}(s \tau+ r \sigma) j_0 S(s \tau+ r \sigma)$ with $j_0$ a constant matrix give a curve with a complete square factor.
If we compute the curve with respect to the $\tau$ factorization (taking the worldsheet signature to be Minkowskian) we find that
\begin{align}
g(z)=(r+s z)^2(a+b z+c z^2) ,\quad a,b,c\in \mathbb{C},
\end{align}
where $a,b,c$ depend on $r,s$.
We could have chosen some other combination of $\tau$ and $\sigma$ for the factorization
(a more natural choice could be to rotate the worldsheet coordinates so that the solution depends only on $\tau$ and then compute the curve),
but, we would get the same form for the curve.
See section \ref{sec:completeFactorizationEx} for more details.

Curves which correspond to completely factorized solutions which also satisfy the Virasoro constraints take the form
\begin{align}
g(z)=(1-z^2)(a+b z)^2,\quad
\mathrm{or}
\quad
g(z)=(1-z^2)(1\pm z)(a+b z),\quad a,b\in \mathbb{C}.
\end{align}
The first form corresponds to the null cusp solution and solutions which are related to it by a worldsheet $\mathrm{SO}(2)$ rotation.
The circular Wilson loop algebraic curve takes the second form, where it might not be obvious at first sight how it is related to completely factorized solutions (see discussion and footnote below (\ref{eq:diffeq2})).

\subsection{Reconstruction of the String Solution}\label{sec:reconstruction}

Given a curve of the form (\ref{eq:AdS3AlgCurve}) we can reconstruct a factorized string solution (up to conformal transformations and worldsheet translations).
One can probably use the reconstruction procedure presented in \cite{Janik:2012ws}, but here just sketch a straightforward procedure independent of the analytical properties which is easy to implement in most cases discussed in this paper.
First we distinguish between two cases\footnote{It can be shown that these two cases are relates to one general solution to $S^{-1}\partial_\tau S=const$, but the solution is quite complicated and in order to get the $c_4=0$ case one has to take the limit carefully. Thus, we find it simpler to distinguish these two cases, although it might not be necessary.}: $c_4\neq 0$ or $c_4= 0$.
If $c_4\neq 0$ we may start with the following ansatz
\begin{align}
j_\sigma=S^{-1}\left(
          \begin{array}{cc}
            \frac{c_3}{4 i \sqrt{c_4}} & \beta(\sigma) \\
            \gamma(\sigma)  & -\frac{c_3}{4 i \sqrt{c_4}} \\
          \end{array}
        \right)S,\quad
j_\tau=
S^{-1}\left(
          \begin{array}{cc}
            \delta(\sigma) & \epsilon(\sigma) \\
            \zeta(\sigma)  & -\delta(\sigma) \\
          \end{array}
        \right)S,\quad
S=e^{\tau \sigma_3 \sqrt{\frac{c_4}{4}}}.
\end{align}
$c_0,c_1,c_2$ of (\ref{eq:AdS3coeff}) fix three of the functions, say $\delta,\epsilon$ and $\zeta$ in terms of $\beta$ and $\gamma$.
Plugging these into the equations of motion and MC equations leaves us with two first order non-linear coupled differentials equation for $\beta(\sigma)$ and $\gamma(\sigma)$,
\begin{align}\label{eq:diffeq1}
\gamma ' = &
\frac{i \left(c_3^3-4 c_2 c_3 c_4+8 c_1 c_4^2-16 c_3 c_4 \beta   \gamma  \right)}{64 c_4^{3/2} \beta }\nonumber\\
&
-\frac{i}{64 c_4^{3/2} \beta  }\bigg(\left(c_3^3-4 c_2 c_3 c_4+8 c_1 c_4^2\right)^2+16 c_4 \beta   \gamma   \left(16 c_2 c_3^2 c_4-3 c_3^4-16 \left(c_2^2+c_1 c_3\right) c_4^2\right.\nonumber\\
&
+\left.64 c_0 c_4^3-16 c_4 \beta   \gamma   \left(8 c_2 c_4-3 c_3^2+16 c_4 \beta   \gamma  \right)\right)\bigg)^{1/2}
\end{align}
\begin{align}\label{eq:diffeq2}
\beta ' = &
-\frac{i \left(c_3^3-4 c_2 c_3 c_4+8 c_1 c_4^2-16 c_3 c_4 \beta   \gamma  \right)}{64 c_4^{3/2} \gamma }\nonumber\\
&
-\frac{i}{64 c_4^{3/2} \gamma}\bigg(\left(c_3^3-4 c_2 c_3 c_4+8 c_1 c_4^2\right)^2+16 c_4 \beta   \gamma   \left(16 c_2 c_3^2 c_4-3 c_3^4-16 \left(c_2^2+c_1 c_3\right) c_4^2\right.\nonumber\\
&
+\left.64 c_0 c_4^3-16 c_4 \beta   \gamma   \left(8 c_2 c_4-3 c_3^2+16 c_4 \beta   \gamma  \right)\right)\bigg)^{1/2}
\end{align}
These equations are generally complicated and we do not give their explicit solution.
For specific curves they can be quite easily solved.
For example, in the case of the circular Wilson loop the curve is given by $(c_0,c_1,c_2,c_3,c_4)=(-\alpha,0,2\alpha,0,-\alpha)$ and we get
\begin{align}
\gamma '-\gamma \sqrt{(\alpha+\beta \gamma )}=0,\quad
\beta '-\beta \sqrt{(\alpha+\beta \gamma )}=0,
\end{align}
with the solution
\begin{align}
\gamma(\sigma)=\gamma_1\beta(\sigma)=\frac{2\alpha e^{\sqrt{\alpha}(\sigma+\gamma_2)}}{1-\gamma_1\alpha e^{2\sqrt{\alpha}(\sigma+\gamma_2)}}.
\end{align}
Different values of $\gamma_1$ and $\gamma_2$ correspond to target space conformal transformations and worldsheet translations respectively\footnote{Note that taking $\gamma_1=e^{-\sqrt{\alpha}\gamma_2}$ and the $\gamma_2\rightarrow -\infty$ limit corresponds to $\gamma(\sigma)=2\alpha e^{\sqrt{\alpha}\sigma}$ and $\beta(\sigma)=0$. This shows how the circular Wilson loop algebraic curve has the form of a completely factorized solution. One really has to go back and find $\epsilon,\zeta,\delta$ in order to really see this property. We note that this transformation is singular in the sense that it is not invertible.}.

After we find the currents we can continue and find the group element $g$ and read the explicit solution.
If $c_0\neq 0$, the group representative takes the form
\begin{align}
g=e^{\tau \sigma_3 \sqrt{\frac{c_0}{4}}}\left(
                         \begin{array}{cc}
                           a(\sigma) & b(\sigma) \\
                           c(\sigma) & d(\sigma) \\
                         \end{array}
                       \right) S(\tau).
\end{align}
Of course there are only three independent functions because of $\det{g}=1$ which eliminates one variable.
Comparing $g^{-1}\partial_{\tau} g\equiv j_\tau$ gives a set algebraic equations which further eliminate two more functions.
Finally, $g^{-1}\partial_{\sigma} g\equiv j_\sigma$ gives one first order differential equation which fixes $g$.
Note that the algebraic curve is invariant under conformal transformations $g\rightarrow l g r$ with $l,r\in \mathrm{SL}(2)$ being constant matrices.

If $c_0=0$, then instead of the $e^{\tau \sigma_3 \sqrt{\frac{c_0}{4}}}$ factor in $g$ we insert $\left(
                                                                                                    \begin{array}{cc}
                                                                                                      1 & 0 \\
                                                                                                      \rho~ \tau & 1 \\
                                                                                                    \end{array}
                                                                                                  \right)
$, $\rho\in \mathbb{C}$.
Similarly, given a curve with $c_4=0$ but $c_3\neq 0$, we cannot use the same $S$ as before, since in that case, $c_4=0$ implies $c_3=0$.
In this case we may use
\begin{align}
j_\sigma=S^{-1}\left(
          \begin{array}{cc}
            \alpha(\sigma) & \beta \\
            \gamma(\sigma)  & -\alpha(\sigma) \\
          \end{array}
        \right)S,\quad
j_\tau=
S^{-1}\left(
          \begin{array}{cc}
            \delta(\sigma) & \epsilon(\sigma) \\
            \zeta(\sigma)  & -\delta(\sigma) \\
          \end{array}
        \right)S,\quad
S=
\left(
\begin{array}{cc}
 1 & 0 \\
 \kappa~\tau & 1 \\
\end{array}
\right).
\end{align}
The rest of the reconstruction follows the same lines as the construction for the $c_4\neq 0$.
In case where $c_4=c_3=0$, we should still use the same $S$ (in this case, the previous $S$ is the special case of $\kappa=0$).

It can be shown the the most general solution to $S^{-1}\partial_\tau S=const.$ is related to the $S$ matrices given above up to conformal transformations.

\section{The AdS$_5$ Case}\label{sec:AdS5S5Curves}

In this section we would like to extend the analysis given above for string solutions in an AdS$_3$ background to solutions in an AdS$_5$ background.
The AdS$_5$ background is a coset manifold $\mathrm{SU}(2,2)/\mathrm{SO}(4,1)$ and not a group manifold.
However, it is still possible to write a sigma model as for the AdS$_3$ background using a special parametrization of the $\mathrm{SU}(2,2)$ group element.
A convenient choice is given in \cite{alday-2006-018,arutyunov-2009}\footnote{In these references, a similar choice for the real form $\mathrm{SU}(4)$ is given which allows one to analyze the $\mathrm{S}^5$ manifold in a similar way.}
\begin{align}\label{eq:coserRep}
g=\left(
\begin{array}{cccc}
 0 & v_3 & v_1 & v_2 \\
 -v_3 & 0 & -v_2^* & v_1^* \\
 -v_1 & v_2^* & 0 & v_3^* \\
 -v_2 & -v_1^* & -v_3^* & 0 \\
\end{array}
\right),
\end{align}
with $|v_1|^2+|v_2|^2-|v_3|^2=-1$.
For global AdS coordinates we use
\begin{align}
v_1=\sinh\rho\cos\psi e^{i\phi},\quad
v_2=\sinh\rho\sin\psi e^{i\varphi},\quad
v_3=\cosh\rho e^{i t},\quad
\end{align}
and for the Poincare patch we use
\begin{align}
v_1=\frac{y^2+x^2-1+2 i x_1}{2 y},\quad
v_2=\frac{x_2+i x_3}{y},\quad
v_3=\frac{y^2+x^2+1+2 i x_0}{2 y},\quad
\end{align}
where $x^2=x^\mu x_\mu$.

The resulting characteristic polynomial equation $\det(L-y)=0$ gives a polynomial of order 4 in $y$.
Due to the fact the $L$ is traceless, it is possible to write the equation in the following form
\begin{align}
\det(L-y)=y^4 -\frac{1}{2} y^2 \Tr(L^2)-\frac{1}{3} y \Tr(L^3)+\frac{1}{8}\left(\Tr(L^2)^2 - 2\Tr(L^4)\right)=0.
\end{align}

As for the AdS$_3$ case we can write in principle the explicit dependence of the curve on the spectral parameter and traces of the currents, but we do not give here the explicit result.
We note that the coefficient of $y^2$ is given by the same function that we had for the AdS$_3$ algebraic curve (\ref{eq:AdS3AlgCurve}), so for example we can immediately detect whether the Virasoro constraints are satisfied according to this function (see section \ref{sec:Virasoro}).
We give an example for a string solution in AdS$_5$ in section \ref{subsec:AdS5ex}.
The analysis is identical for string solutions on S$^5$ where instead of $\mathrm{SU}(2,2)$ we use $\mathrm{SU}(4)$.

\section{$\mathbb{Z}_4$ Super-Coset Models}\label{sec:fac_sol_4_coset_man}
In this section we generalize the analysis given above for sigma-models on group manifolds to sigma-models on $\mathbb{Z}_4$ super-coset spaces \cite{Metsaev:1998it} (see also \cite{Berkovits:1999zq,Adam:2007ws,Zarembo:2010sg}).
Given a solution for a sigma-models on a super-coset space with a factorization of the flat-connection, $A(\tau,\sigma)=S^{-1}(\tau)A_0(\sigma)S(\tau)$ with $j_S=S^{-1}d S=const.$, the construction of the Lax operator identical to the one where the background is a groups manifold, namely (\ref{eq:LaxOp}).
However, a factorization of the MC one-form, $J(\tau,\sigma)=S^{-1}(\tau)J_0(\sigma)S(\tau)$, does not immediately imply a factorization of the flat-connection as in the group manifold case.

Starting with such a factorization for $J$, we define
\begin{align}
J=J_e+J_o\equiv S^{-1} J_0 S=S^{-1} (J_e)_0 S+S^{-1} (J_o)_0 S,
\end{align}
where $J_e$ and $J_o$ are the even and odd components of $J$ respectively.
The $\mathbb{Z}_4$ decomposition is given by
\begin{align}
J^{(0/2)}=\frac{1}{2}(1\pm\Omega)J_e,\quad
J^{(1/3)}=\frac{1}{2}(1\mp i\Omega)J_o,
\end{align}
with $\Omega^2 J=(-)^{|J|}J$, where $|J|=0$ if $J$ is even and $|J|=1$ if $J$ is odd.
Requiring that $J^{(i)}$ will also decompose as $J^{(i)}(\tau,\sigma)=S^{-1}(\tau)J^{(i)}_0(\sigma)S(\tau)$ implies that
\begin{align}
[j_S-\Omega(j_S),\Omega(J)]=0.
\end{align}
This means that $j_S\in \mathcal{H}_0$, or equivalently $S\in H$, where $H\subset G$ is the invariant locus of the $\mathbb{Z}_4$ automorphism.

The common parameterizations of the super-coset representative $g$
(e.g. $g=\exp(x\cdot P)y^D$ for Poincare coordinates where $P$ and $D$ are the translation and dilatation generators respectively)
are non-linear and the factorization property might be obscured.
Thus, it is more practical to use the parameterizations introduced in section \ref{sec:AdS5S5Curves}.

\section{Examples}\label{sec:examp_4_group_man}
In this section we give several examples for applications of the procedure described in previous sections for constructing the algebraic curve and for reconstructing the solution from a given curve.
Some of the examples were studied in previous papers (\cite{Janik:2012ws,Ryang:2012uf}) using the method of \cite{Janik:2012ws}.
Most of the examples are for string solutions in $\mathrm{AdS}_3$, some of them are good solutions satisfying the Virasoro constraints, while other which do not satisfy the Virasoro constraints require extra excitations on the sphere.

\subsection{Generalization of the $q\bar q$ Potential Solution}
In this section we study a generalization of the $q\bar q$ potential solution in the sense of starting with a general string solution of the form
\begin{align}
x_1&=\kappa \tau,\nonumber\\
x_2&=F(\sigma),\nonumber\\
y&=G(\sigma),
\end{align}
where we use a Euclidean worldsheet metric as well as a Euclidean AdS$_3$ space\footnote{Note that this is not the generalized $q\bar q$ potential introduced in \cite{Drukker:2011za} and discussed in \cite{Janik:2012ws} in the context of the algebraic curve.}.
Taking the group element to be of the form
\begin{align}
g=\left(
          \begin{array}{cc}
            \frac{i x_1 +x_2}{y^2} & \frac{1}{y} \\
             -\frac{x_1^2 +x_2^2+y^2}{y}& \frac{i x_1 -x_2}{y^2} \\
          \end{array}
        \right),\quad
\end{align}
the solution admits a factorization with
\begin{align}
S(\tau)=\left(
          \begin{array}{cc}
            1 & 0 \\
            i \kappa \tau & 1 \\
          \end{array}
        \right),\quad
j_S=\left(
          \begin{array}{cc}
            0 & 0 \\
            i \kappa  & 0 \\
          \end{array}
        \right).
\end{align}
If we assume that $F(\sigma)$ and $G(\sigma)$ solve the equations of motion, we have
\begin{align}
& \Tr(j_\tau j_\sigma) = 0\nonumber\\
& \Tr(j_\tau j_\tau-j_\sigma j_\sigma) = \frac{2 \left(\kappa ^2-F'^2-G'^2\right)}{G^2} \equiv v =const.
\end{align}
Eliminating $F'$ in favor of $v$ and plugging the MC one-form in the algebraic curve equation (\ref{eq:AlgCurveEq})
we find
\begin{align}
y^2=-\det(A_\tau-j_S)
=
\frac{z}{2(z^2-1)^2}\left(
(z^2-1)\frac{\kappa\sqrt{4\kappa^2-2v G^2-4 G'^2}}{G^2}
-v z
\right),
\end{align}
where $\frac{\kappa\sqrt{4\kappa^2-2v G^2-4 G'^2}}{G^2}$ must be a constant for any solution.
So at this point we already have a pretty general form for the curve without having to solve any differential equation and find the explicit solution.

For completeness, we solve one differential equation for $G$ which gives
\begin{align}
G(\sigma)
=
\frac{\kappa }{\alpha  \beta }\mathrm{cn}\left(\beta (\sigma +c),1-\alpha ^2\right)
\end{align}
where $\alpha,\beta,c$ are constants, and $v=2\beta^2(2\alpha^2-1)$, and $\mathrm{cn}$ is the Jacobi elliptic function.
Plugging the solution to the algebraic curve equation gives
\begin{align}
y^2=
-\frac{z \left(\left(1-z^2\right) \alpha  \sqrt{1-\alpha ^2}+z \left(1-2 \alpha ^2\right)\right) \beta ^2}{\left(1-z^2\right)^2}.
\end{align}

We can further solve the equation for $F$ which gives
\begin{align}
F(\sigma)=
&a+ b \sigma
-
\kappa  \sqrt{\left(1-\alpha ^2\right)}
\Bigg(\frac{\alpha(c+\sigma )}{1-\alpha ^2}\nonumber\\
&
-
\frac{\text{E}\left(\text{am}\left(\beta  (c+\sigma ),1-\alpha ^2\right),1-\alpha ^2\right)
\left(\alpha ^2+\left(1-\alpha ^2\right) \text{cn}\left(\beta  (c+\sigma ),1-\alpha ^2\right)^2\right)}{\alpha\beta\left(1-\alpha ^2\right) \text{dn}\left(\beta  (c+\sigma ),1-\alpha ^2\right) \sqrt{1-\left(1-\alpha ^2\right) \text{sn}\left(\beta  (c+\sigma ),1-\alpha ^2\right)^2}}\Bigg)
\end{align}
where $a,b$ are constants, am is the Jacobi amplitude and sn and dn are also standard Jacobi elliptic functions.

Let us note several limits of the solution.
The first is the one for which the Virasoro constraints are satisfied, that is $\alpha=1/\sqrt{2}$ \cite{Chu:2009qt} (which was discussed in \cite{Janik:2012ws}).
If we also take $\beta=1$ and $c=0$ then we recover the known $q\bar q$ solution with
\begin{align}
G(\sigma)
=
\sqrt{2}~\kappa ~\mathrm{cn}\left(\sigma,1/2\right).
\end{align}
We get another simple solution when $\alpha=1$, $\beta=1$ and $c=0$ where
\begin{align}
G(\sigma)
=
\kappa \cos \sigma,\quad
F(\sigma)=const.,
\end{align}
which describes a straight folded string attached to the boundary, moving along a line (that is, the limit where the quark and anti-quark are at the same point).
The limit $\alpha\rightarrow 0$ with $\beta=1$ and $c=0$ gives $G\rightarrow \frac{1}{\alpha \cosh\sigma}+\mathcal{O}(\alpha)$.

Other solutions with $0<\alpha<1$ describe the $q\bar q$ pair moving along a line with different distance between them (where the distance goes to infinity when $\alpha\rightarrow 0$), but these solutions should be supplemented by motion on the sphere since they do not satisfy the Virasoro constraints.

Another rather trivial limit is $\kappa\rightarrow 0$. This leaves us with the real solution given by
\begin{align}
G(\sigma)
=
a \mathrm{sech}(\beta (\sigma + c)),\quad
F(\sigma)
=
b+a \tanh(\beta (\sigma +c)),
\end{align}
with
$v=-2\beta^2$ and the curve $g(z)=-\beta^2 z^2$.
This is a static open string with the endpoints attached to the boundary.

\subsection{Two Point Correlation Function}
In this example we study the closed string solution given in \cite{Janik:2010gc} describing a two point Correlation function of operator which are dual to classical spinning strings.
We use the the form of the solution before the special conformal transformation, given by
\begin{align}
x_1&=\tanh\rho_0 \cos(\omega\tau+\sigma)e^{\kappa\tau},\nonumber\\
x_2&=\tanh\rho_0 \sin(\omega\tau+\sigma)e^{\kappa\tau},\nonumber\\
y&=\cosh^{-1}\rho_0 e^{\kappa\tau},
\end{align}
where $\kappa^2+\omega^2=1$ and $\rho_0=const$.
The worldsheet metric is Lorentzian and the solution is embedded in a Euclidean AdS$_3$.
In this case we note that we can extract an algebraic curve equation by factorizing the flat connection as $A(\tau,\sigma)=S(\tau)A_0(\sigma)S^{-1}(\tau)$
or $A(\tau,\sigma)=R(\sigma)\tilde A_0(\tau)R^{-1}(\sigma)$.
The matrices $S$ and $R$ take the form $S=e^{(\kappa-i\omega)\tau/2}$ and $R=e^{-i\tau/2}$ respectively.
This is what we call a \emph{completely factorized solution}.

The algebraic curve associated with the $S$ factorization is given by
\begin{align}
y^2=-\det(A_\tau-j_S)
=
\frac{(z-i \kappa +\omega )^2 \left(-1+z^2 (\kappa -i \omega )^2+2 z (i \kappa +\omega ) \cosh{2\rho_0}\right)}{4 \left(-1+z^2\right)^2}
\end{align}
One can see the complete square factor $(z-i \kappa +\omega )^2$ which implies that the solution is complectly factorized (see section \ref{sec:completeFactorizationEx}).
The algebraic curve associated with the $R$ factorization is given by
\begin{align}
y^2=-\det(A_\sigma-j_R)
=
&
\frac{(z+i \kappa + \omega  )^2 \left(-z^2+(\kappa +i \omega )^2+2 z (-i \kappa +\omega ) \cosh{2\rho_0}\right)}{4 \left(-1+z^2\right)^2 }.
\end{align}

These two curves look different, since the solution is not symmetric with respect to $\tau$ and $\sigma$.
The curves obviously coincide when $\kappa=0,\omega=1$ since then the solution depends on $\tau$ and $\sigma$ in the same way.
The curve then takes the form
\begin{align}
y^2=-\det(A_\sigma-j_R)
=
-\frac{1+z^2-2 z \cosh{2 \rho_0}}{4 (1-z)^2}.
\end{align}

The energy-momentum tensor components are
\begin{align}
&\Tr(j_\sigma j_\tau)=2\omega\sinh^2\rho_0,\nonumber\\
&\Tr(j_\tau j_\tau+j_\sigma j_\sigma)=2\cosh{2\rho_0}+\kappa^2-\omega^2-1,
\end{align}
so generally the Virasoro constraints are not satisfied.

\subsection{Completely Factorized Solutions}\label{sec:completeFactorizationEx}
In the previous example we found that the solution was factorized both with respect to $\tau$ and $\sigma$.
This is also the case for the null cusp solution \cite{Kruczenski:2002fb}.
Here we consider the most general completely factorized solutions and their properties.
We use Euclidean worldsheet for the general analysis.

We start by considering solutions of the form
$j(\tau,\sigma)=T^{-1}(\tau,\sigma) j_0 T(\tau,\sigma)$, where $T(\tau,\sigma)=S(\tau)R(\sigma)=\exp(\sigma_3(s\tau +r\sigma))$, $r,s\in \mathbb{C}$.
We plug this current into the equations of motion and get a set of simple algebraic equations which can be easily solved. The result is
\begin{align}
(j_0)_\sigma=\left(
\begin{array}{cc}
 \alpha  & \beta  \\
 \gamma  & -\alpha  \\
\end{array}
\right),\quad
(j_0)_\tau=
\left(
\begin{array}{cc}
 \frac{r^2+s^2-r \alpha }{s} & -\frac{ r \beta }{s} \\
 -\frac{ r \gamma }{s} & -\frac{r^2+s^2-r \alpha }{s} \\
\end{array}
\right)
\end{align}
\begin{align}
\Tr(j_\tau j_\tau-j_\sigma j_\sigma)=&-2 \alpha ^2+\frac{2 \left(r^2+s^2-r \alpha \right)^2}{s^2}-2 \beta  \gamma +\frac{2 r^2 \beta  \gamma }{s^2},\nonumber\\
\Tr(j_\tau j_\sigma)=&\frac{2 \alpha  \left(r^2+s^2-r \alpha \right)-2 r \beta  \gamma }{s},
\end{align}
\begin{align}
y^2=\frac{(s+i r z)^2 \left((r+i s z-\alpha )^2+\beta  \gamma \right)}{s^2 \left(-1+z^2\right)^2},
\end{align}
where $\alpha,\beta,\gamma\in \mathbb{C}$.
We see that for such solutions the curve has a complete square factor.

In order for the Virasoro constraints to be satisfied we can take $\beta=\frac{\alpha  \left(r^2+s^2-r \alpha \right)}{r \gamma }$ and either $r=\pm i s$ or $\alpha=r$
which gives the curves
\begin{align}
y^2=-\frac{s(1+z) (s+s z+2 i \alpha )}{\left(1-z^2\right)}\quad
\mathrm{and}
\quad
y^2=\frac{(s+i r z)^2}{1-z^2},
\end{align}
respectively.
The second curve corresponds to the null cusp solution with $(s,r)=\frac{1}{\sqrt{2}}(1,1)$, and its 2d worldsheet rotations \cite{Kruczenski:2007cy,Roiban:2007dq}.
The folded string in the scaling limit yields a curve with the $(s,r)=(i\kappa,0)$ curve\footnote{
Note that when computing the curve with respect to the $\tau$ factrization the null cusp's curve of \cite{Kruczenski:2002fb} is different then the folded string which is related by an $\mathrm{SO}(2,4)$ target space transformation, a Wick rotation and 2d worldsheet rotations \cite{Kruczenski:2007cy}. The reason is that our procedure is sensitive to the worldsheet transformations. This result is different then the one presented in \cite{Ryang:2012uf} where both curves are same. In the completely factorized solutions there is an ambiguity regarding the choice of the factorization matrix , and different choices for the factorization matrices could yield similar results for the curves.}.
All these string solutions live on the surface $x_1^2-x_2^2=\frac{1}{2} y^2$.

The first curve gives a real solution for Minkowskian worldsheet.
These strings live on the surface given by $x_1^2-x_2^2=-\frac{(s-\alpha )^2}{4 s \alpha } y^2$ embedded in a Euclidean AdS$_3$ space.
These solutions depends only on $(\sigma+\tau)$ so they describe infinite "static" strings on this surface.

Let us comment that algebraic curves which are associated with completely factorized solutions by having a complete square factor may give non-trivial string solutions which are not completely factorized, but are related to such solutions by conformal transformation and a worldsheet translation which are singular (in the sense that they are not invertible).
One example which we described before is the circular Wilson loop.
Another example is an open string solution living on the surface of an $\mathrm{S}^2$ embedded in a Euclidean AdS$_3$ space, where the center of the sphere is on the boundary of the AdS space and the sphere is contracting. This solution does not satisfy the Virasoro constraints by itself (unless $n=1$).
The solution is given by
\begin{align}
x_1&=A e^{-\sqrt{n}\tau}(n\cos\sigma+\cos(n\sigma)),\nonumber\\
x_2&=A e^{-\sqrt{n}\tau}(n\sin\sigma-\sin(n\sigma)),\nonumber\\
y  &=2 A \sqrt{n} e^{-\sqrt{n}\tau}\sin\left(\frac{(1+n)\sigma}{2}\right).
\end{align}
The energy-momentum tensor components are given by
\begin{align}
\Tr(j_\tau j_\tau-j_\sigma j_\sigma)&=\frac{1}{2} (n-1)^2,\nonumber\\
\Tr(j_\tau j_\sigma)&=0,
\end{align}
and the algebraic curve is given by
\begin{align}
y^2=\frac{\left(\sqrt{n}+z\right)^2 \left(1-\sqrt{n} z\right)^2}{4 \left(1-z^2\right)^2}.
\end{align}
As for the circular Wilson loop, the way to get the completely factorized solution is to make a
conformal transformation by $g\rightarrow g e^{i \xi n/2}$ followed by a translation $\sigma\rightarrow \sigma+\xi$
and taking the limit $\xi\rightarrow -i\infty$.
Note that $\sigma$ ranges from $0$ to $\frac{2\pi}{1+n}$, otherwise the strings continues to $-y$.
The projection of the solution on the $x_1$-$x_2$ plane is a spiky contracting string with $n+1$ spikes (if $\sigma\in [0,2\pi]$).
We will introduce a similar solution where the string spins in section \ref{subsec:AdS5ex}.

In case where $c_4=0$ we use $T(\tau,\sigma)=\exp(\frac{\sigma_1-i\sigma_2}{2}(s\tau +r\sigma))$ and generally get the curve
\begin{align}
y^2=\frac{(s+i r z)^2 \left(\alpha ^2+\beta  (\gamma-r-i s z )\right)}{s^2 \left(1-z^2\right)^2}
\end{align}
again with the complete square factor.
The Virasoro constraints are given by
\begin{align}
\Tr(j_\tau j_\tau-j_\sigma j_\sigma)=&-\frac{2 \left(r^3 \beta +r s^2 \beta -r^2 \left(\alpha ^2+\beta  \gamma \right)+s^2 \left(\alpha ^2+\beta  \gamma \right)\right)}{s^2},\nonumber\\
\Tr(j_\tau j_\sigma)=&\frac{r^2 \beta +s^2 \beta -2 r \left(\alpha ^2+\beta  \gamma \right)}{s}.
\end{align}
In order to solve the Virasoro constraints we take $\frac{2 r \alpha ^2}{r^2+s^2-2 r \gamma }$ and either $\alpha=0$ which gives us the trivial curve $y^2=0$ or $r=\pm i s$ which gives
\begin{align}
y^2=\frac{i s (1-z)^2 (1+z) \alpha ^2}{\gamma (1-z)^2}
\end{align}

\subsection{Folded Spinning String in AdS$_3$}
The folded spinning string solution in a Minkowskian AdS$_3$ space in global coordinates is given by \cite{Gubser:2002tv}
\begin{align}
t=&\kappa \tau,\nonumber\\
\phi=&\omega \tau,\nonumber\\
\rho=&\rho(\sigma),\nonumber\\
\end{align}
The solution to the equations of motion gives
\begin{align}
\rho(\sigma)=\pm i~\mathrm{am}\left((\sigma+\gamma_2)(\omega^2-\kappa^2-\gamma_1)^{1/2}\bigg|\frac{\kappa ^2-\omega ^2}{\kappa ^2-\omega ^2+\gamma_1}\right).
\end{align}
This solution does not satisfy the Virasoro constraint and is not periodic for the general integration constants $\gamma_1,\gamma_2$.
It is easy to check that this solution is factorized with $S=\exp(i\tau(\kappa-\omega)/2)$ if we use
\begin{align}
g=\left(
    \begin{array}{cc}
      e^{i t}\cosh\rho & e^{i \phi}\sinh\rho \\
      e^{-i \phi}\sinh\rho & e^{-i t}\cosh\rho \\
    \end{array}
  \right)
.
\end{align}
Thus, we easily compute the algebraic curve equation
\begin{align}
y^2=-\frac{\left(\kappa -z^2 \kappa +\omega +z^2 \omega \right)^2-4 z^2 \gamma_1}{4 \left(1-z^2\right)^2}.
\end{align}
The Virasoro constraints are satisfied for $\gamma_1=\omega^2$ where we have
\begin{align}
y^2=-\frac{\left(1-z^2\right)((\kappa + \omega )^2-(\kappa -\omega )^{2 }z^2)}{4 \left(1-z^2\right)^2}.
\end{align}
We note that if $\kappa=\omega$ (as in the long string scaling limit) we retrieve the complete factorized solution
\begin{align}
y^2=-\frac{4\kappa^2\left(1-z^2\right)}{4 \left(1-z^2\right)^2},
\end{align}
which is related to the null cusp solution by 2d worldsheet transformations followed by a conformal transformation and a Wick rotation \cite{Kruczenski:2007cy}, as discussed above (the algebraic curve in this limit was also discussed in \cite{Ryang:2012uf}).
We also note that when $\kappa=0$ or $\omega=0$ we get the circular Wilson loop algebraic curve
\begin{align}
y^2=-\frac{\kappa^{2 }\left(1-z^2\right)^2}{4 \left(1-z^2\right)^2}.
\end{align}
This must be the case since in this limit the solutions coincide.

\subsection{$\left<W(C)\Tr z^J\right>$ Correlation Function}\label{sec:WrtZ^Jexample}
The algebraic curve for the $\left<W(C)\Tr z^J\right>$ correlation function was worked out in \cite{Janik:2012ws}.
For this solution the factorization is given with respect to $\sigma$ coordinate on the worldsheet.
The solution is given by \cite{Zarembo:2002ph}
\begin{align}
x_1 &= \frac{\sqrt{1+j^2}e^{j\tau}}{\cosh(\sqrt{1+j^2}\tau+\xi)}\cos\sigma,\nonumber\\
x_2 &= \frac{\sqrt{1+j^2}e^{j\tau}}{\cosh(\sqrt{1+j^2}\tau+\xi)}\sin\sigma,\nonumber\\
y &= (\sqrt{1+j^2}\tanh(\sqrt{1+j^2}\tau+\xi)-j)e^{j\tau},\nonumber\\
\xi &= \log(j+\sqrt{1+j^2}),
\end{align}
where the worldsheet is Euclidean and the target space is a Euclidean $\mathrm{AdS}_3$ space and $j=J/\sqrt{\lambda}$.
The factorization matrix is given by $S=\exp(i\sigma_3 \sigma/2)$.
Plugging this into the Lax operator $L=A_\sigma-j_S$ yields the curve
\begin{align}
y^2=-\frac{\left(1-2 j z-z^2\right)^2}{4 \left(1-z^2\right)^2}.
\end{align}
This solution does not satisfy the Virasoro constraints which are given by
\begin{align}
\Tr(j_\tau j_\tau-j_\sigma j_\sigma)&=2j,\nonumber\\
\Tr(j_\tau j_\sigma)&=0.
\end{align}

\subsection{Open Strings Solution in AdS$_5$}\label{subsec:AdS5ex}
Up to now we have considered examples of string solutions in AdS$_3$ space.
In the following example we show how the procedure can be applied for string solutions in AdS$_5$ (or more precisely Euclidean AdS$_4$).
We give a non-trivial factorized solution to the equations of motions in AdS$_5$.
The solution describes an open string attached the AdS boundary that moves constantly along one direction. The string has a fixed profile and spins with time around the $y$ axis in the $x_1$-$x_2$ plane.
This solution is a bit similar to the $q \bar q$ potential, but where the pair of quarks also rotate.
We use a Euclidean worldsheet and a Minowskian AdS$_5$ space as the target space.

The solution is given by
\begin{align}
x_0&=0,\nonumber\\
x_1&=A \left(n \text{cos}\left(\sigma -\sqrt{n} \tau \right)+\text{cos}\left(n \sigma +\sqrt{n} \tau \right)\right),\nonumber\\
x_2&=A \left(-n \text{sin}\left(\sigma -\sqrt{n} \tau \right)+\text{sin}\left(n \sigma +\sqrt{n} \tau \right)\right),\nonumber\\
x_3&=\kappa \tau,\nonumber\\
y&=\frac{2 \left(A^2 n (1+n)^2+\kappa ^2\right)^{1/2}}{(1+n)}\sin\left(\frac{1}{2} (1+n) \sigma\right),
\end{align}
so the $x_1,x_2,y$ components describe the ellipsoid $x_1^2+x_2^2+\frac{A^2 n (1+n)^2}{A^2 n (1+n)^2+\kappa ^2}y^2=A^2 (1+n)^2$.
For $\kappa=0$, the string profile coincides with the contracting open string described in section \ref{sec:completeFactorizationEx}.
As in that case, we need to limit the range of the string to be $\sigma\in[0,\frac{2\pi}{1+n}]$, or else we would get a negative value for $y$.
Plugging the solution into the coset representative given in (\ref{eq:coserRep}) and computing the MC one-form, we find the factorization $j(\tau,\sigma)=S^{-1}(\tau)j_0(\sigma)S(\tau)$ with
\begin{align}
S=\exp{
\left(\tau
\left(
\begin{array}{cccc}
 0 & \frac{1}{2} i \left(\sqrt{n}-\kappa \right) & -\frac{i \kappa }{2} & 0 \\
 \frac{1}{2} i \left(\sqrt{n}-\kappa \right) & 0 & 0 & \frac{i \kappa }{2} \\
 \frac{i \kappa }{2} & 0 & 0 & -\frac{1}{2} i \left(\sqrt{n}+\kappa \right) \\
 0 & -\frac{i \kappa }{2} & -\frac{1}{2} i \left(\sqrt{n}+\kappa \right) & 0 \\
\end{array}
\right)
\right)
},
\end{align}
so the form of $j_S$ is just the matrix in the exponent's argument.

We do not give the explicit forms of $j$ and $A$ since they are lengthy and not illuminating.
the resulting algebraic curve equation is given by
\begin{align}
&y^4+
y^2
\frac{A^2 n (1+n)^2 \left((n-1) z+i \sqrt{n} \left(1+z^2\right)\right)^2-\left((1+n (4+n)) z^2-n \left(1+z^4\right)\right) \kappa ^2}{2 \left(1-z^2\right)^2 \left(A^2 n (1+n)^2+\kappa ^2\right)}
\nonumber\\
&
+\frac{
A^4 n^2 (1+n)^4 \left(z-n z-i \sqrt{n} \left(1+z^2\right)\right)^4
+\kappa ^4\left(n+z^2\right)^2 \left(1+n z^2\right)^2
}
{16 \left(1-z^2\right)^4 \left(A^2 n (1+n)^2+\kappa ^2\right)^2}\nonumber\\
&
-\frac{
2 A^2 \kappa ^2 n (1+n)^2 \left(n-(1+n (4+n)) z^2+n z^4\right) \left(z-n z-i \sqrt{n} \left(1+z^2\right)\right)^2
}
{16 \left(1-z^2\right)^4 \left(A^2 n (1+n)^2+\kappa ^2\right)^2}
=0.
\end{align}
In the limit where $\kappa=0$ the solution is embedded in Euclidean AdS$_3$ where the string lives on $\mathrm{S}^2$ and the algebraic curve becomes
\begin{align}
y^2=\frac{\left(\sqrt{n}+i z\right)^2 \left(i+\sqrt{n} z\right)^2}{4 \left(1-z^2\right)^2}.
\end{align}
The energy-momentum tensor components are given by
\begin{align}
&\Tr(j_\sigma j_\tau)=-\frac{2 A^2 (n-1) n^{3/2} (1+n)^2}{A^2 n (1+n)^2+\kappa ^2},\nonumber\\
&\Tr(j_\tau j_\tau-j_\sigma j_\sigma)=\frac{(1+n)^2 \left(A^2 n (1+(n-6) n)+\kappa ^2\right)}{A^2 n (1+n)^2+\kappa ^2}.
\end{align}

The Virasoro constraints are satisfied only for $n=1$ with $A=\kappa/2$.
The second constraint can be solved by $A=\frac{\kappa }{\sqrt{-n+6 n^2-n^3}}$ where there are real solutions up to $n=3+2\sqrt{2}$ (or $n=5$ for integers).
In the $n=1$ and $A=\kappa/2$ case the algebraic curve equations reduced to
\begin{align}
y^4+\frac{y^2}{2}+\frac{\left(1+z^2\right)^2}{16 \left(1-z^2\right)^2}=0,
\end{align}
so that
\begin{align}
y^2=\frac{(i\pm z)^2}{4 \left(1-z^2\right)},
\end{align}
which looks like two copies of null cusp curves.

\section{Discussion}\label{sec:discussion}

In this paper we showed how an algebraic curve can be easily constructed for factorized string solutions, by introducing a simple Lax operator.
The Lax operator is not unique (and so also the curve),
but nonetheless we give a prescription for constructing it in an unambiguous way for any factorized string solution, without having to introduce arbitrary functions.
We study the properties of the curve, especially for string solutions on AdS$_3$ background, and give examples for applications of our procedure for various string solutions.
Our procedure yields similar results to those found in \cite{Janik:2012ws} and \cite{Ryang:2012uf} up to a rescaling of the curve's equation and 2d worldsheet rotations.

The algebraic curve that we study here is different from the algebraic curve originally introduced in \cite{Kazakov:2004qf}, where the quasi-momenta properties depend also on global properties of the solution, such as the energy and other charges.
The curves that we study are local in this sense and do not depend on this data (integration is not required in the process).
This is expected since we are also able to find curves for open string solutions which should break some of the global symmetries.
Also, in order to extract the algebraic curve from the derivative of the quasi-momenta as in the original construction one has to
perform a birational transformation \cite{Dorey:2006zj}, while in \cite{Janik:2012ws} and in this paper one does not make such transformations.
It is interesting to understand better the relation between the resulting curves from these two procedures.

Since the curve is not associated with the global charges it is not clear if there exists a generating function for an infinite set of conserved charges like the monodromy matrix.
Such a function exists for a class of integrable open string solutions \cite{Dekel:2011ja}, but this function is also defined globally, and moreover is not compatible with the open string ending on the boundary of AdS.
For the factorized solutions one can naively construct many functions of the flat-connection which are time independent, since the trace of any power of the flat-connection is conserved by definition.
Thus, for example one can take $\Tr (e^{A_\alpha})$ (note that integration is not necessary as it is for a general solution) and then expand the function in terms of the spectral parameter.
However, it is not guaranteed that the infinitely many "charges" (the coefficients in the expansion) are independent.
It is interesting find out if it is possible to come out with such a function for these string solutions.

All the worked out examples for constructing an algebraic curve for string solutions with no non-contactable loops fall into the class of factorized solutions.
It is interesting to find out if there are algebraic curves associated with non-factorized solutions (that do not have a monodromy) and study their structure.
We also note that any curve of the form introduced here (that is, any genus-1 curves) correspond to factorized solutions.

Our analysis of factorized string solutions on AdS$_5$ background is less detailed then the one for solutions on AdS$_3$ background.
It is interesting to further study the curve's structure on this background and more generally the $\mathrm{AdS}_5\times \mathrm{S}^5$  and $\mathrm{AdS}_4\times \mathbb{C}\mathrm{P}^3$ supercoset backgrounds.

\section*{Acknowledgements}
I would like to thank
T. Bargheer,
T. Klose,
J. Minahan,
Y. Oz,
and
K. Zarembo
for valuable discussions.

\bibliography{postdoc_V3}
\bibliographystyle{JHEP}
\end{document}